\documentclass{vgtc}                          

\usepackage{mathptmx}
\usepackage{graphicx}
\usepackage{times}

\usepackage{latexsym}
\usepackage{epsfig}
\usepackage{psboxit}
\usepackage{wrapfig}
\usepackage{subfigure}
\usepackage{algorithm}
\usepackage{algorithmic}
\usepackage{amsfonts}
\usepackage{graphicx}

\onlineid{59}

\vgtccategory{Research}

\vgtcinsertpkg

\title{An Eye Tracking Study into the Effects of Graph Layout}

 \author{Weidong Huang\thanks{e-mail: weidong.huang@nicta.com.au}}
 \affiliation{\scriptsize School of Information Technologies, University of Sydney, Australia \\ IMAGEN Program, National ICT Australia Ltd.}

\bigskip



\abstract{Graphs are typically visualized as node-link diagrams.
Although there is a fair amount of research focusing on crossing
minimization to improve readability, little attention has been paid
on how to handle crossings when they are an essential part of the
final visualizations. This requires us to understand how people read
graphs and how crossings affect reading performance.

As an initial step to this end, a preliminary eye tracking
experiment was conducted. The specific purpose of this experiment
was to test the effects of crossing angles and geometric-path
tendency on eye movements and performance. Sixteen subjects
performed both path search and node locating tasks with six
drawings. The results showed that small angles can slow down and
trigger extra eye movements, causing delays for path search tasks,
whereas crossings have little impact on node locating tasks.
Geometric-path tendency indicates that a path between two nodes can
become harder to follow when many branches of the path go toward the
target node. The insights obtained are discussed with a view to
further confirmation in future work.
} 

\keywords{eye tracking, edge crossing, geometric path, evaluation,
graph drawing}

\CRcatlist{ 
\CRcat{H.1.2}{Models and Principles}{User/Machine Systems}{Human
Factors};
  \CRcat{H.5.0}{Information Interfaces and Presentation}{User
Interfaces}{Evaluation/methodology}
}

\begin{document}



\maketitle

\bigskip
\bigskip

\section{Introduction} \label{sec:intro}

Graphs are typically visualized as \emph{node-link diagrams}. A
graph can be drawn in many different ways by simply changing the
layout of nodes. A growing number of empirical studies have shown
that graph layout affects not only readability, but also the
understanding of the underlying data. In particular, \emph{edge
crossings} (or link crossings) has long been a major concern in
graph drawing; it is commonly accepted and employed as a general
rule that the number of crossings should be reduced as much as
possible~\cite{kau}. However, in practice, \emph{crossing
minimization} is a hard problem in designing algorithms for graph
drawing~\cite{garey}. There are also many graphs in which crossings
are not removable. Although there is a fair amount of research
focusing on crossing minimization (e.g.,~\cite{eades,junger,mutzel})
in the literature, little attention has been paid on how to handle
crossings when they are an essential part of the final
visualizations.

\begin{figure}[t]

  \centering
  \subfigure[]{
     \label{fig:ex1}
     \includegraphics[height=0.9in]{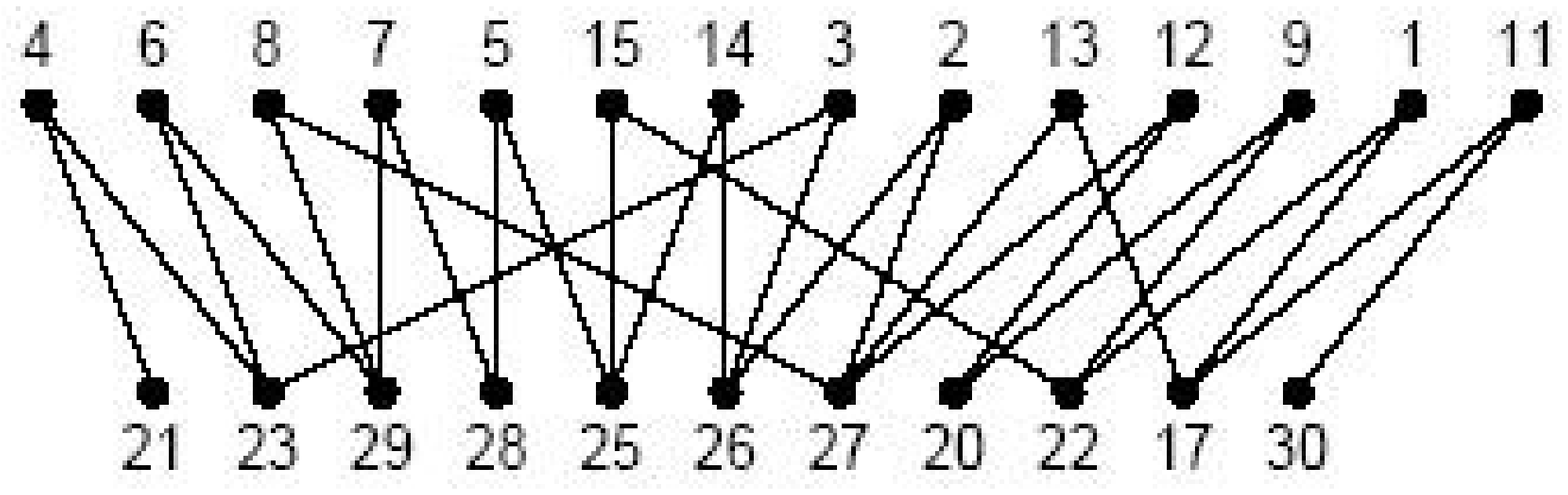}}
  \hspace{0.5in}
  \subfigure[]{
     \label{fig:ex2}
     \includegraphics[height=0.9in]{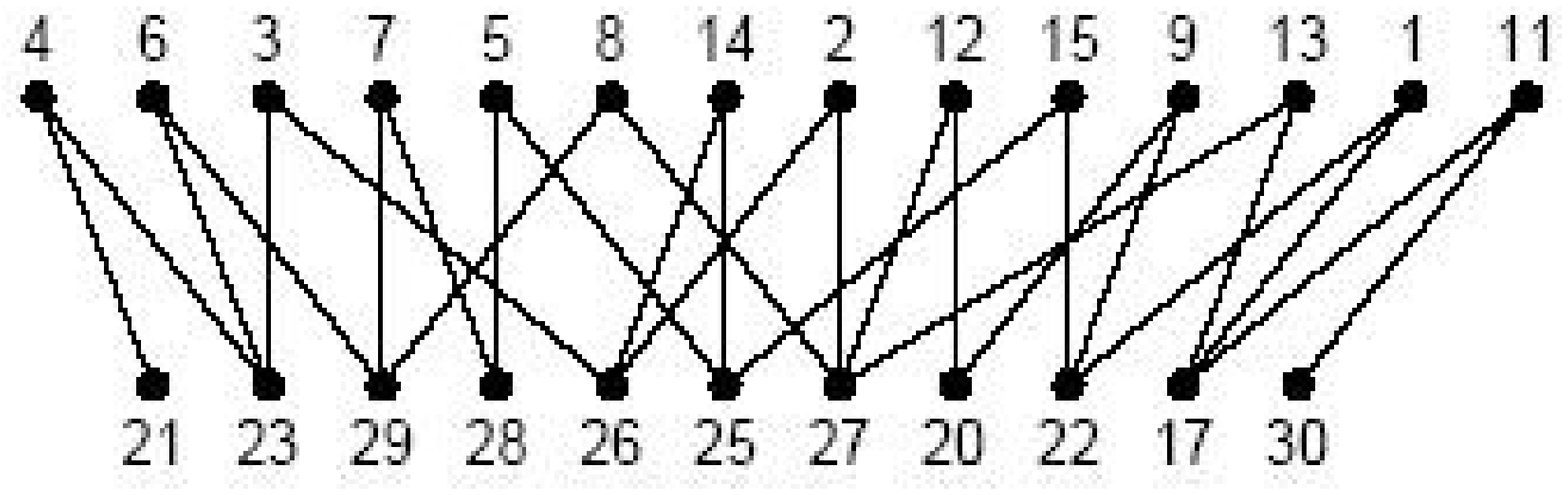}}
  \caption{Two drawings of the same graph. (a) $k$-planarization drawing, (b) minimal-crossing-number drawing. Adopted from~\cite[Figure 2]{mutzel}}
  \label{fig:ex}
\end{figure}

Some researchers have pointed out that different crossing styles may
have different degrees of impact. Take the two drawings in
Figure~\ref{fig:ex}, as an example. These two drawings were of a
graph and drawn using two different approaches:
$k$-planarization~\cite{mutzel} and
minimal-crossing-number~\cite{junger}, respectively. The drawing in
Figure~\ref{fig:ex1} has 34 crossings, which is $41\%$ more
crossings than the drawing in Figure~\ref{fig:ex2} has (24
crossings). However, as indicated in~\cite{mutzel}, an informal
evaluation revealed that the former drawing was considered as having
less crossings and being more readable. Further, not only the
collective crossing pattern has a role in affecting graph
perception, but also the individual crossing angle. For example, as
mentioned in~\cite{hu05, ware02}, when edges cross at
nearly-90-degree angles, they are less likely to be confusing than
when crossing at acute angles.

In addition, more and more empirical studies are available showing
that in some situations, crossings may not be as bad as we normally
think (e.g.,~\cite{hu05, hu06}). For example, in perceiving
sociograms (node-link diagrams for social networks), it was found
that crossings are important only for tasks that involve path
tracing~\cite{hu06}. Even when sociograms are drawn to convey
specific information, such as how many social groups there are in
the network, it is more desirable to cross edges connecting the
group members~\cite{hu06}. It is also possible that drawing graphs
without crossings can make some structural features less apparent,
such as symmetry.

Thus, when the cost of crossing reduction cannot be justified, or
when crossings become unavoidable, the questions arise: How can we
reduce the negative impact of crossings to the minimum? In what
situations can we simply ignore the presence of crossings, or even
make use of them? To answer these questions, we need to have
knowledge of how and when crossings, or visual layouts in a broader
sense, affect graph understanding. In addition, it is also essential
for us to have a good understanding of how people read graphs.

\bigskip
\subsection{Related Work}

User studies investigating layout effects can be divided into two
groups according to the graphs used: abstract graphs and domain
graphs (such as sociograms, UML diagrams).

Purchase~\cite{pu97} conducted a user study examining the effects of
five graph drawing aesthetics (symmetry, edge crossings, angular
resolution, and orthogonality) on task performance. It was
demonstrated that minimizing crossings was overwhelmingly beneficial
in understanding graph structure; edge crossings was \lq\lq by far
the most important aesthetic\rq\rq ~compared to the other four. In
an experiment that was to examine several aesthetics within the same
set of computer-generated diagrams, Ware et al.~\cite{ware02} found
that good path-continuity was also a positive factor for path search
tasks. They also demonstrated that for shortest path tasks, \lq\lq
it is the number of edges that cross the shortest path itself that
is important, rather than the total number of edge crossings in the
drawing\rq\rq~\cite{ware02}.

Korner et al.~\cite{korner} investigated the effects of visual
properties of hierarchical graphs on task response speed: planarity
(edge crossing), slopes (edge orientation), and levels (hierarchy).
Analysis of response latencies showed that crossings was the most
influential variable. \lq\lq It is the general disarrangement
present in crossed drawings that causes the slower comprehension
speed~\cite{korner}\rq\rq, no matter whether the graph elements in
question are involved with crossings or not. This is quite different
from what was found in~\cite{ware02} on the effects of crossings. We
will come back to this matter later in this paper.

In investigating layout effects on sociogram perception, McGrath et
al.~\cite{mc97} administered a user study. Five different drawings
of a network were used. In each of these drawings, the spatial
arrangement varied in Euclidean distance between two nodes and nodes
to the center of the drawing. Subjects were asked to perform domain
specific tasks. It was found that both network structure and spatial
arrangement of nodes influenced the understanding of network
structural features~\cite{mc97}. In another study, McGrath et
al.~\cite{mc96} found that the perception of network groups can be
significantly affected by the visual clusters appearing in the
sociogram.

\bigskip

\subsection{Motivation} \label{sec:related}

Many observations in graph layout evaluations are mainly based on
task response time and accuracy. This approach tells us what the
consequences will be in terms of task performance when a particular
layout is to be used. However, this approach treats the human mind
as a \lq\lq black box\rq\rq, and therefore cannot explain where the
time is spent and how the accuracy is affected.

\begin{figure}[h]

\centering
\includegraphics[height=1.7in]{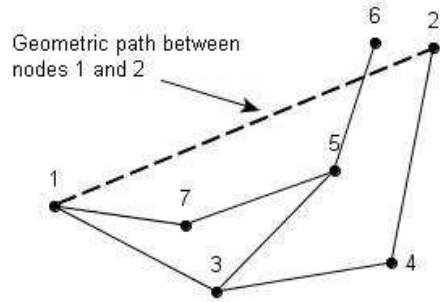}
\small \caption{Illustration of geometric-path tendency. Note that
the dashed line is not part of the graph. To find the path between
nodes 1 and 2, if search always starts from node 1, people tend to
follow the path 1-7-5-6 first, the path 1-3-5-6 second, and the path
1-3-4-2 third. } \label{fig:path}
\end{figure}

As an initial attempt to understand how people read graphs, an eye
tracking study was conducted in~\cite{hu05} to understand how
crossings affect task performance. Subjects were given five pairs of
crossing and non-crossing drawings and asked to find a shortest path
between two given nodes for each drawing. Their response times and
eye movements while performing the tasks were recorded. It was
reported that only one crossing drawing took the subjects a
significantly longer time than the corresponding non-crossing
drawing. The video analysis showed that:

\begin{enumerate}

\item  Crossings had little impact on subjects' eye
movements; it appeared that those crossings were simply ignored by
the subjects during path searching.

\item  It was the edges going towards the target node that
distracted eyes and caused delays. In other words, as illustrated in
Figure~\ref{fig:path}, in performing the shortest path task,
subjects tended to follow edges which were close to the geometric
path of the two nodes. If the edges were not part of the shortest
path, they had to go back and start searching again, which took time
and caused errors. For simplicity, this graph reading behavior is
termed as \emph{geometric-path tendency}.

\end{enumerate}

This was quite surprising since it indicated that crossings were not
the major time consuming factor as expected. However, a closer look
at the study revealed the following facts:
\begin{enumerate}

 \item  The graphs used in~\cite{hu05} were sparse, and small with the largest graph
containing only 11 nodes and 15 edges.

\item  The crossing angles in the crossing drawings were generally quite large (nearly 90 degrees).

\end{enumerate}

When graphs are small and crossing angles are large, the impact of
crossings may be too small to be significant. The user study
described in this paper was aimed to address the above points. It
was hoped that this experiment could provide some useful insights
for future design of more formal experiments investigating layout
effects.

\bigskip
\subsection{Outline}
The rest of the paper is organized as follows. Section 2 presents
the experiment, followed by the results in section 3. The findings
obtained in this experiment are discussed in section 4. Finally,
section 5 concludes the paper with an outline of future work.

\bigskip
\bigskip

\section{Experiment} \label{sec:experiment}

This experiment was conducted to: 1) see how crossing angles affect
eye movements; 2) replicate geometric-path tendency. The first is a
drawing property and the second is a graph reading behavior.

\bigskip
\subsection{Subjects} \label{sec:subjects}

Sixteen subjects were recruited on a completely voluntary basis. All
of them had normal vision and were regular computer users. They had
different degrees of familiarity with node-link diagrams; two of
them had no knowledge at all at the time of participation. The
subjects were reimbursed \$20 each for their time and effort upon
the completion of their tasks.

\bigskip
\subsection{Apparatus}

The testing room contained one operator PC on which an eye tracking
system was running, one subject IBM T41 laptop on which stimulus
diagrams were to be shown, and adjustable chairs and tables.
Adjustments were made to maintain the subject\rq s eyes at a
distance of approximately 40~cm from the 14-inch monitor of the
laptop. The eye tracking system used in the experiment was iViewX
with Headmounted Eye-tracking Device (HED) (SensoMotoric Instruments
GmbH (SMI)). The HED is a helmet to be worn by the subject that
contains an eye camera.

A calibration tool called WinCal was used for visualizing
calibration points and run on the subject laptop, so that the
subject can calibrate while sitting in front of the laptop. The
laptop and the operator PC were connected by a serial line for this
purpose. Once enabled, WinCal can be triggered by the commands from
the operator PC, maximize and minimize itself automatically at the
start and end of calibration, respectively. The calibration area in
iViewX had been set to match the resolution of the laptop monitor,
that is, 1024 $\times$ 768 pixels.

The eye tracker tracks eye movements by observing the position of
the pupil and corneal reflex from the right eye. The system had been
reconfigured so that the content of the laptop monitor screen with
eye gaze position indicated by a gaze cursor can be recorded into
MPEG video files for offline analysis.

\bigskip
\subsection{Stimuli}

Six drawings were grouped into two three-drawing sets: Set 1 for
testing crossing angles, and Set 2 for testing geometric-path
tendency. As can be seen from Table~\ref{tbl:drawings}, relatively
larger graphs were used compared to those in~\cite{hu05}. Note that
for Set 1 drawings in Table~\ref{tbl:drawings}, only some of the
nodes were labeled, though in the real tests, all labels were
visible.

The three drawings (c1, c2 and c3) from Set 1 were of a graph
containing 32 nodes and 43 edges. The graph had 2 components: path
component and condition component. In producing the three drawings,
the layout of the path component remained unchanged. The layout of
the other component was modified to make the three conditions: no
crossings on the path (c1), nearly-90-degree crossings on the path
(c2) and small-angle crossings on the path (c3). c1 was the control
condition that was to compare how eye movements changed when
crossings were introduced in c2 and c3.

In drawing graphs, reducing crossings normally causes a path less
continuous~\cite{ware02}. However, path continuity was also
identified as a notable factor affecting graph
perception~\cite{ware02}. By using two-component graphs and keeping
the path component the same, this confounding effect can be removed.
In addition, since the main purpose for Set 1 drawings was to test
crossing angles, not to test how easy or difficult it is to find the
correct path, the path component contained only one path to avoid
any other confounding effects introduced by multiple paths and
branches. However, subjects were not made aware of these facts
beforehand, although they might have come to realize them after they
had finished the tasks.

The three drawings (f1, f2 and f3) from Set 2 were of another graph
containing 20 nodes and 32 edges. In f1, the shortest path between
nodes 1 and 2 (1-11-15-2) was far away from the geometric path of
nodes 1 and 2 and had no crossings, while the shortest path in f2
(1-6-11-2) and f3 (1-22-11-2) had three crossings (with
nearly-90-degree angles) and was near the geometric path. In
addition, there were more crossings in total in f2 than in f3.

\begin{table*}
\begin{center}

\begin{tabular}{|c ||c |}
\hline Set 1 \,\, & Set 2  \\
\hline
\includegraphics[width=2.2in]{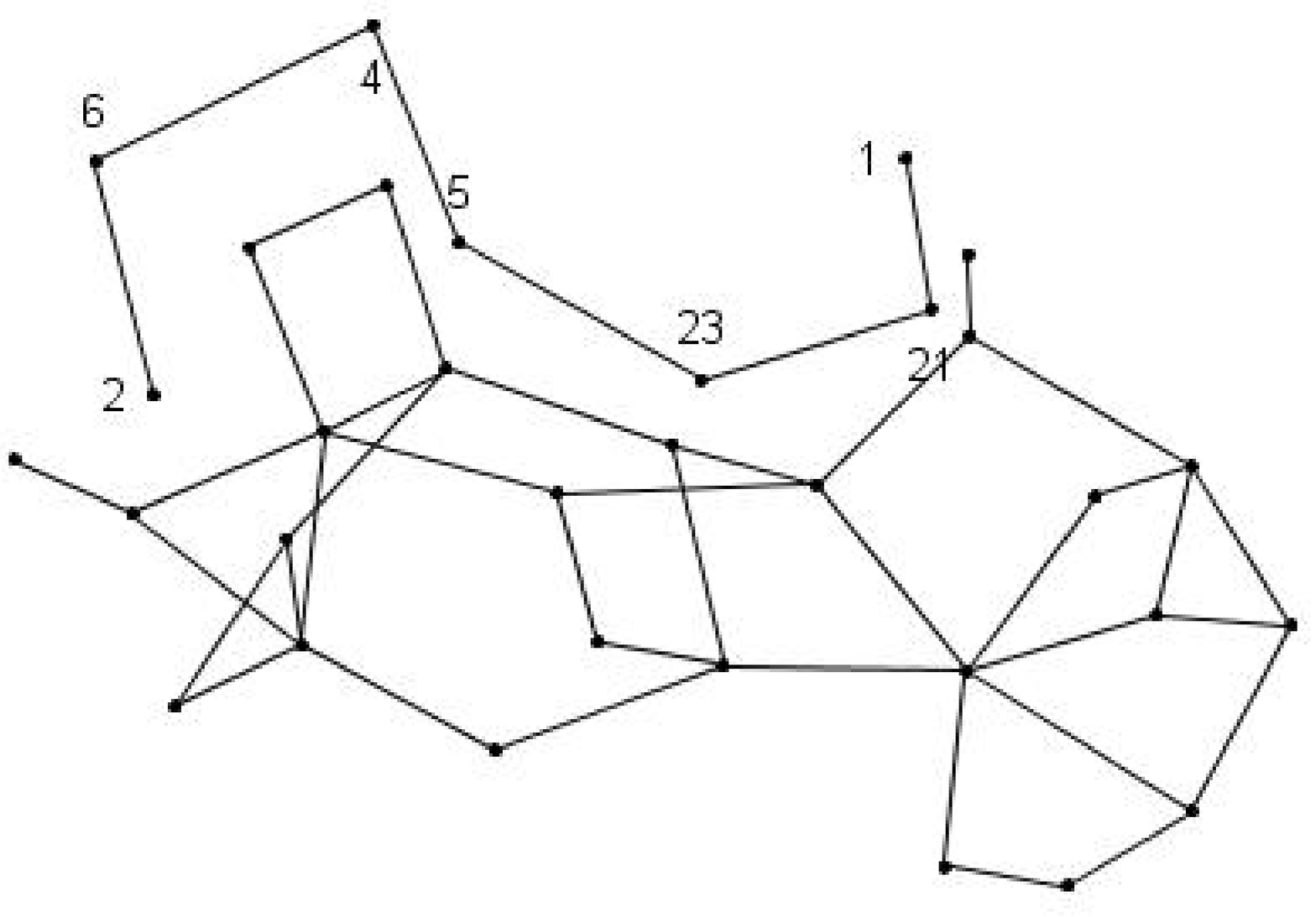} & \includegraphics[width=2.2in]{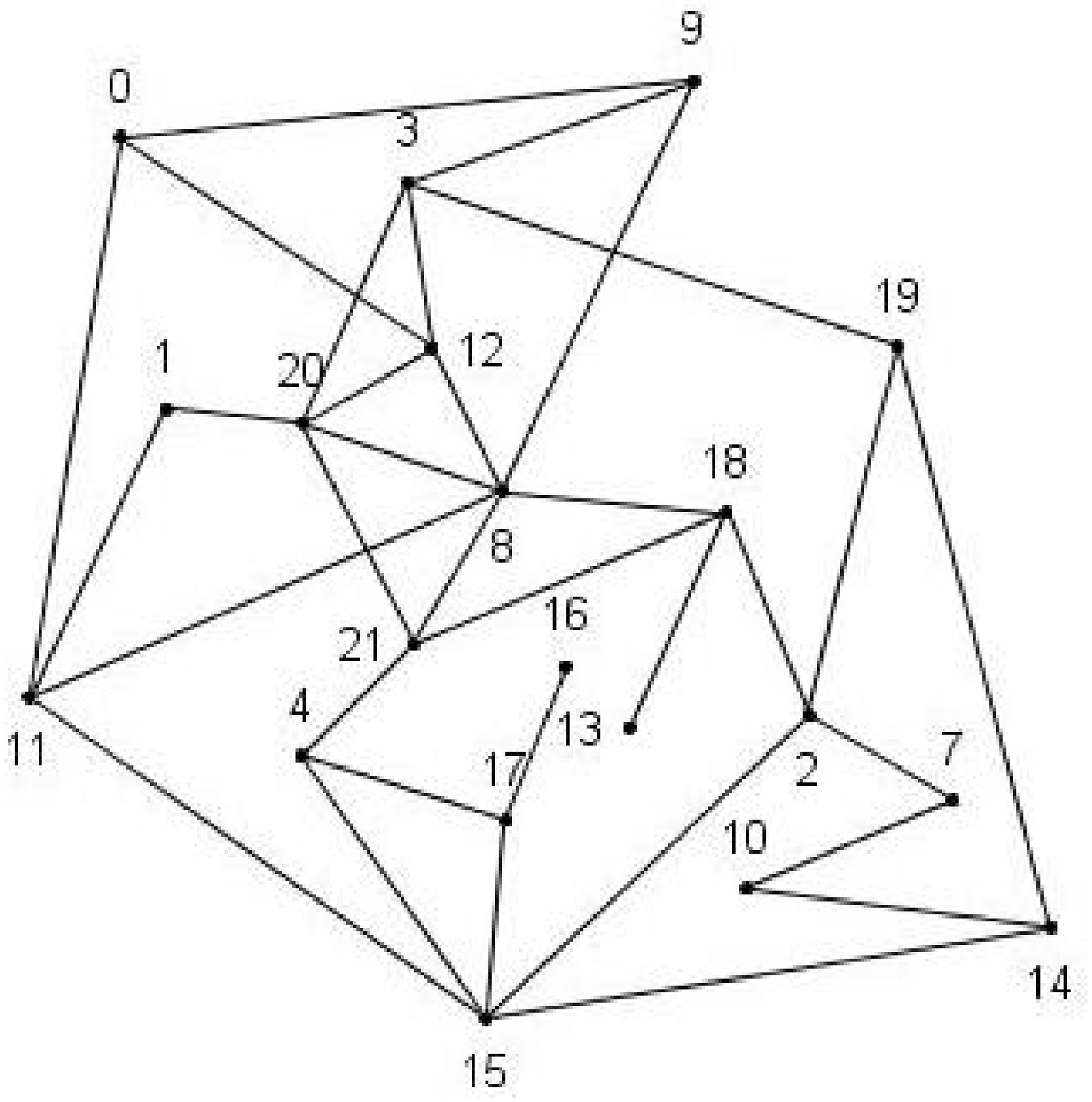} \\
&\\
c1 & f1\\
\hline
\includegraphics[width=2.2in]{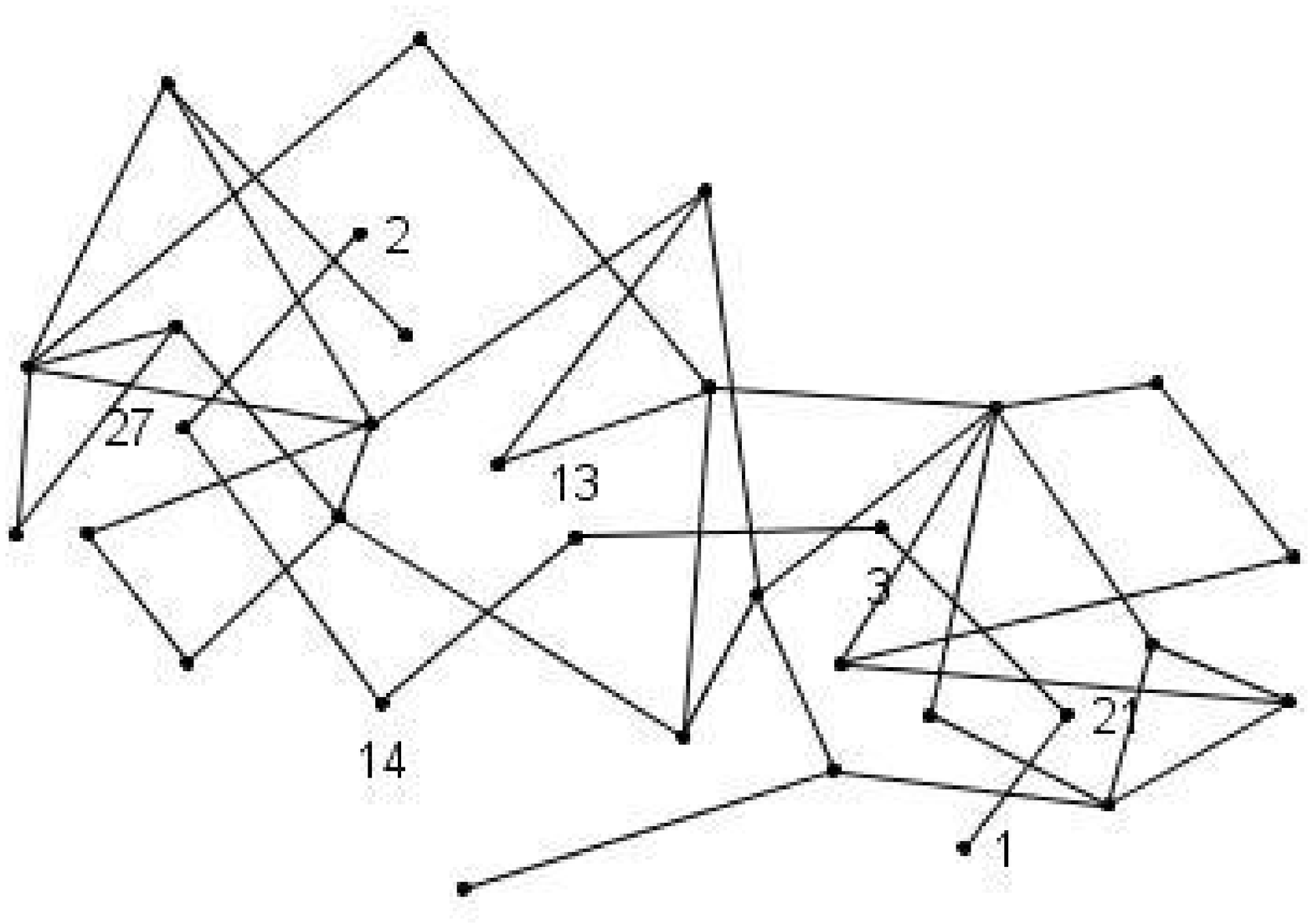} & \includegraphics[width=2.2in]{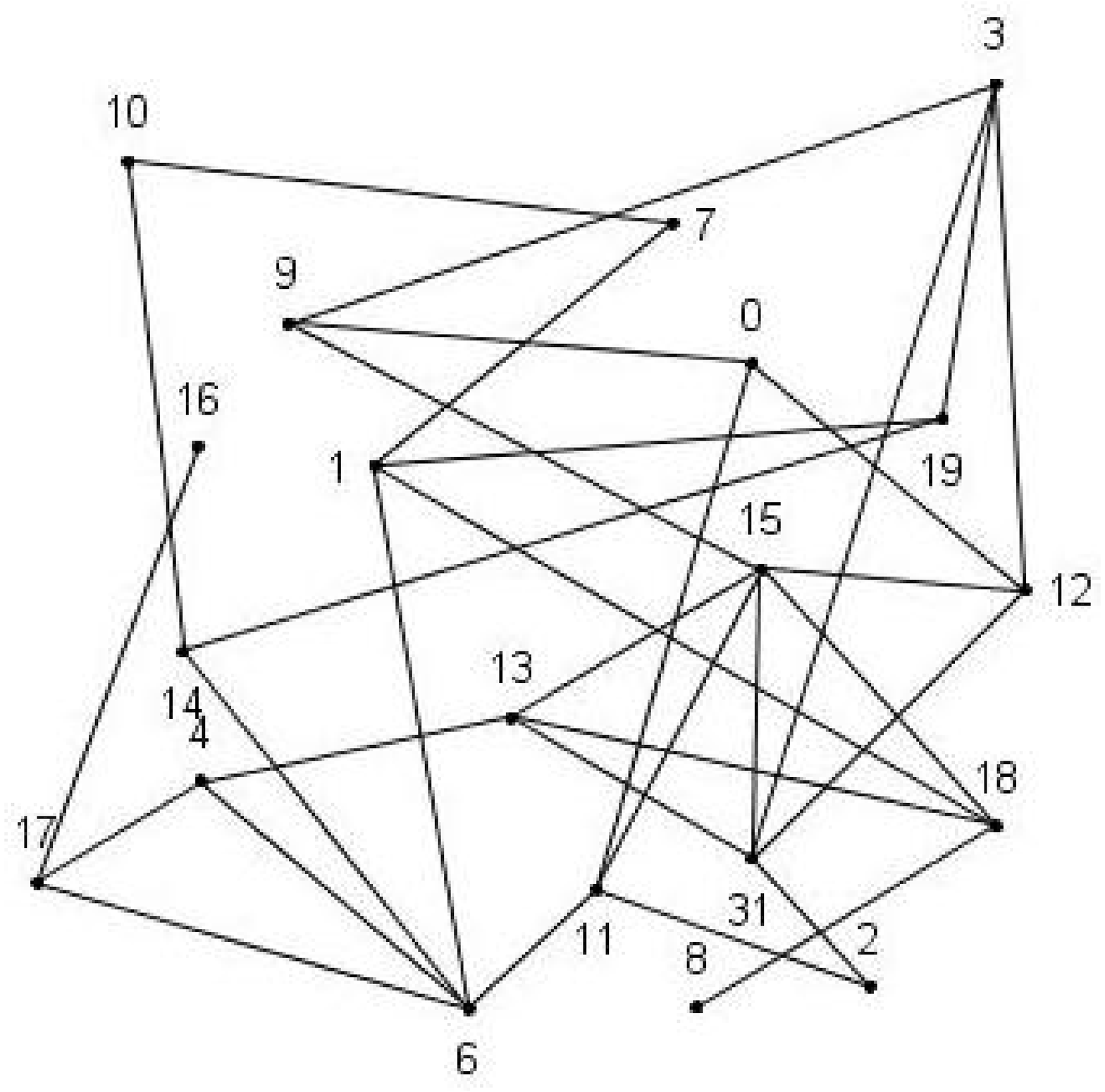} \\
&\\
c2 & f2\\
\hline
\includegraphics[width=2.2in]{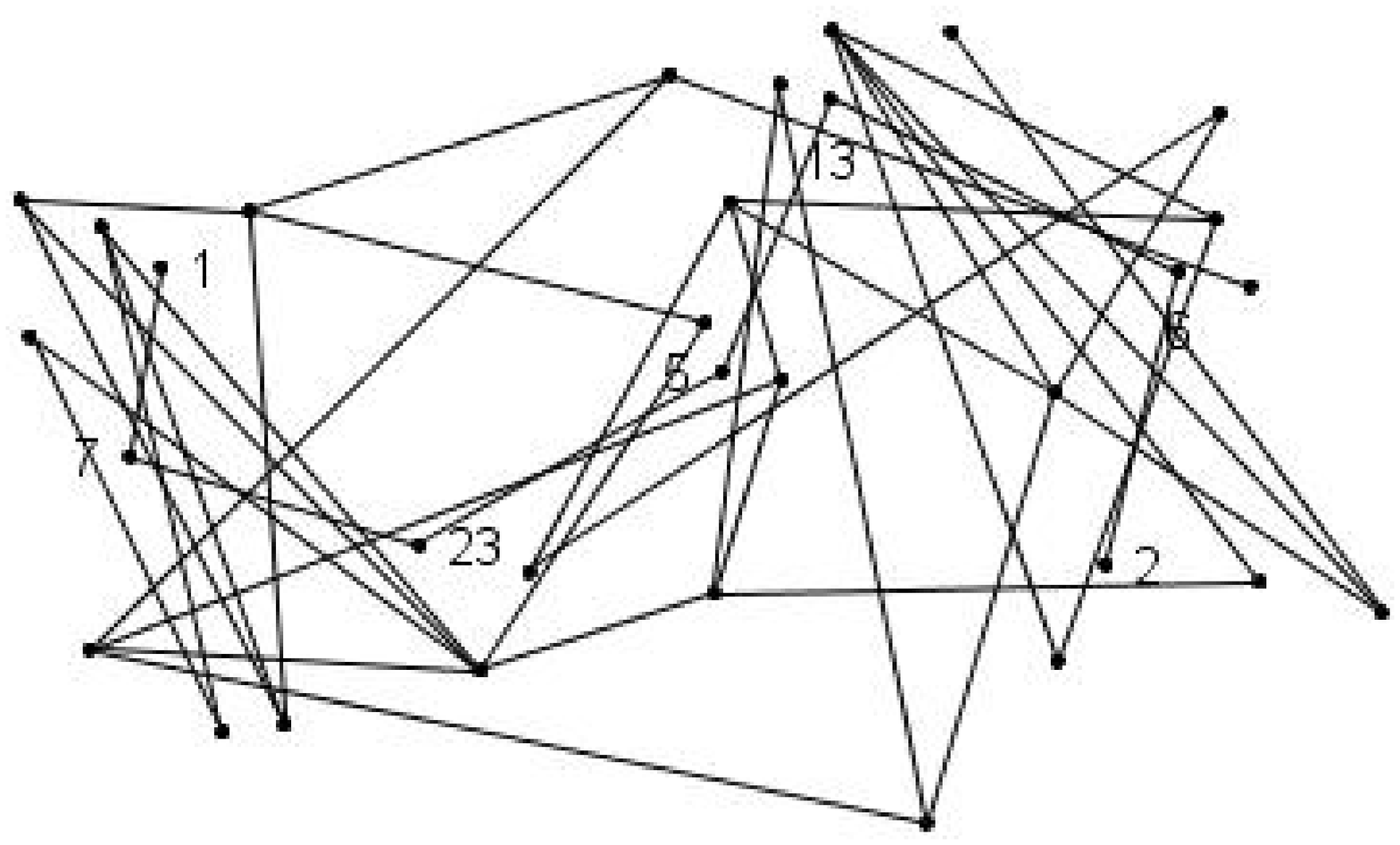} & \includegraphics[width=2.2in]{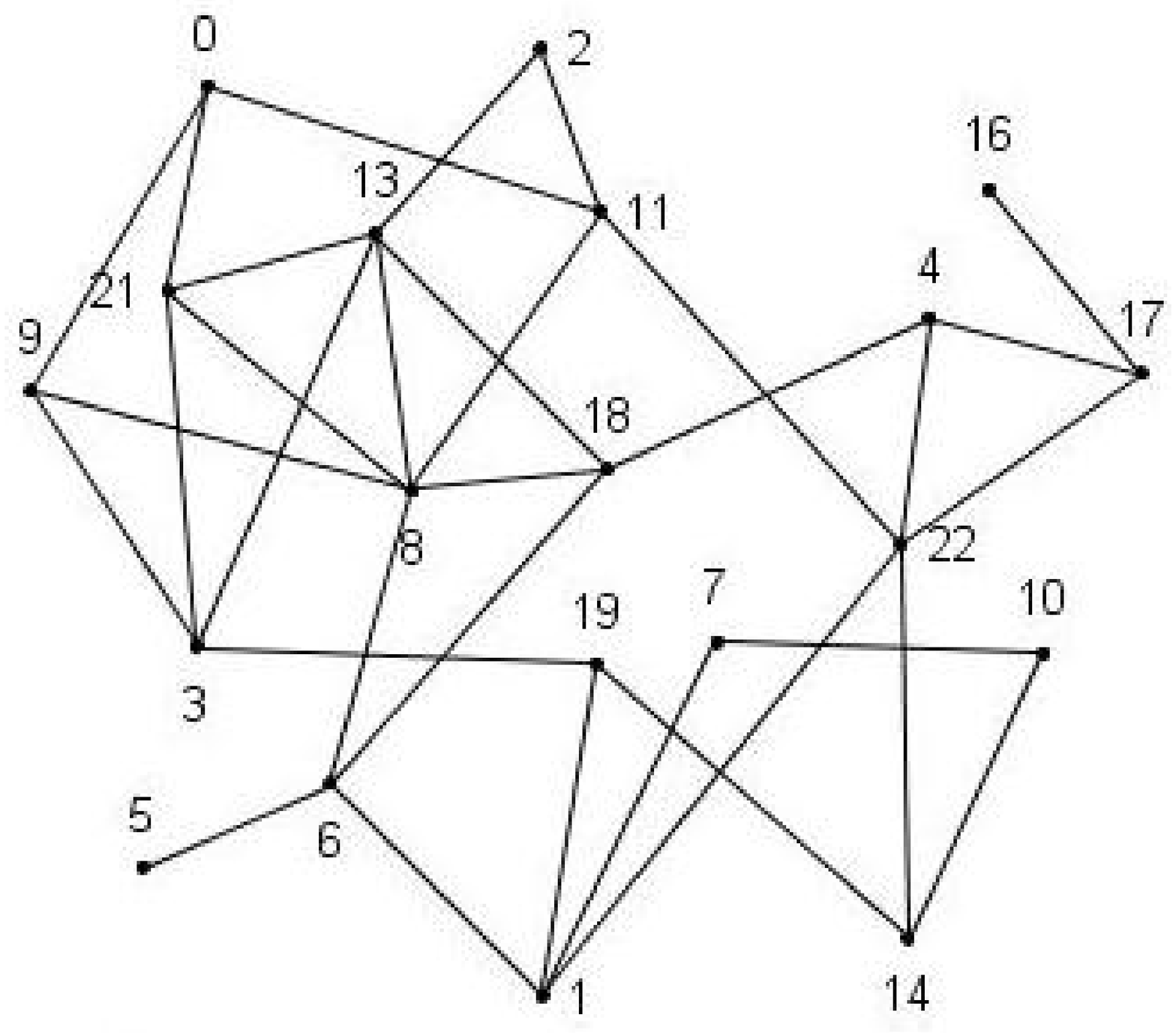} \\
&\\
c3 & f3\\
\hline
\end{tabular}
\caption{Two sets of drawings used in the study}
\label{tbl:drawings}
\end{center}
\end{table*}

\bigskip
\subsection{Tasks} \label{sec:tasks}

Shortest path tasks are typically used in previous studies in
testing the effects of crossings. To understand the effects fully, a
node locating task was also included in this study. After all, among
the seven generic tasks described in ~\cite{gh04}, node locating is
one of the important components in graph perception.

In testing path search tasks, related nodes are normally
pre-selected and highlighted, so that locating the nodes can be
excluded from path searching. As mentioned in
section~\ref{sec:related}, the study~\cite{korner} had a different
finding on the effects of crossings, compared to the finding from
the study in~\cite{ware02}. This might be due to the fact that the
nodes to be considered were not highlighted beforehand
in~\cite{korner}; to find the path between them, the two nodes had
to be located first. Korner et al.~\cite{korner} suspected that
\lq\lq crossings themselves may affect early stages of visual
information processing\rq\rq. \lq\lq Such salient properties
(crossings) are processed in precedence, and draw attention and
distract the visual system from the message of the drawing\rq\rq. If
this is the case, they are likely to happen during the node locating
stage. To see what is really going on when node locating is part of
a path search task, the shortest path task without highlighting
nodes first was also included as one of the tasks.

As such, the following three tasks were used:

\begin{enumerate}
 \item Path task: find the shortest path between nodes 1 and 2. Nodes
 1 and 2 were highlighted.

 \item Node+Path task: find the shortest path between nodes 1 and 2.
 Nodes 1 and 2 were not highlighted.

  \item Node task: find the most connected node.
\end{enumerate}

The first task is a pure path search task, the third is a target
locating task, and the second is the combination of path search and
target locating.

\bigskip
\subsection{Online Task Setting}

A system had been developed for the subject to perform the tasks
online. The system displayed a question first. The subject pressed
the button on the screen, the question disappeared and the
corresponding drawing was then shown. The subject answered the
question by clicking one of the buttons above the drawing; each
button showed one of the possible answers. Once the button was
clicked, a new question was shown, and so on.

Subjects' responses (time and accuracy) were logged by the online
system. Their eye movements were recorded by the eye tracking system
in real time.

The experiment included three sessions; one session for each task.
The order of the three tasks was random. In each session, the
subject had to perform the task with each of the six drawings. The
drawings in each session were displayed randomly. There was a break
between sessions. Just before each session a calibration was
conducted.

Each time when a drawing was shown, the nodes were labeled
differently to avoid possible recognition of the same graph.

\bigskip
\subsection{Procedure} \label{sec:procedure}

All the subjects were given time to read tutorial material, ask
questions and practice. They were also instructed to answer each
question as quickly as possible without compromising accuracy, and
not to use a mouse to help.

The experiment was conducted on an individual basis. After some
practice, subjects performed the tasks online. A post-task
questionnaire was given, and a short interview held with the
subject, following the experiment. Seven of the subjects were chosen
to explain their eye-movement behaviors while watching their own eye
movement videos. The whole experiment took about 50 minutes.

\bigskip
\bigskip

\section{Results}
\bigskip

\subsection{Quantitative Results} \label{sec:results}

Although eye movement data was of the main interest for this study,
the performance data was also analyzed. Given the small number of
subjects and the limited variety of the stimuli used in the study,
quantitative results are presented as additional evidence in support
of eye movement findings, which are described in
sub-section~\ref{sec:video}.

\subsubsection{Response Time}

\begin{table}[h]
\begin{center}
\begin{tabular}{|l| r r r |r r r |}
\hline Drawing ID   & c1 & c2 & c3 & f1 & f2 & f3 \\
\hline\hline Node task & 16.37 & 19.97 & 23.22 & 16.54 & 17.52 & 16.20 \\
\hline Path task & 6.81 & 14.74 & 29.41 & 13.61 & 16.07 & 13.33 \\
\hline \small{Node+Path task} & 9.54 & 16.28 & 33.58 & 15.72 & 15.12 & 21.91 \\
\hline
\end{tabular}

\caption{Median times (sec.) for all the drawings and tasks}
\label{tbl:time}

\end{center}
\end{table}

A non-parametric test of Friedman was used for statistical analysis.
The median times for all responses are shown in
Table~\ref{tbl:time}.

 \textbf{Node Task:} Among Set 1 drawings, the shortest time was spent with c1,
followed by c2, then c3. The test indicated that there were
significant differences in response times ($p = 0.047$). Pairwise
comparisons found that only the difference between c1 and c3 was
significant ($p = 0.006$). For Set 2 drawings, a slightly shorter
time was spent with f3 than with f1. The longest time was spent with
f2. However, the test did not find any significant differences ($p =
0.144$).

\textbf{Path Task:} From Table~\ref{tbl:time}, it can be seen that
the time spent with c2 and c3 was longer than with c1; the longest
time was spent with c3. The test revealed that these differences
were statistically significant ($p < 0.001$). Pairwise comparisons
indicated that the time difference for each pair was also
statistically significant (for each pair, $p \leq 0.001$). For Set 2
drawings, the subjects spent the longest time with f2, followed by
f1. The shortest time was spent with f3. However, the test showed
that these differences were not statistically significant ($p =
0.570$).

\textbf{Node+Path Task:} For Set 1 drawings, the longest time was
spent with c3, followed by c2, then c1. The analysis revealed that
there were significant differences in response time among the three
drawings ($p < 0.001$). Pairwise comparisons indicated that the
difference for each pair was also statistically significant ($p <
0.05$). For Set 2 drawings, the shortest time was spent with f2,
followed by f1. f3 took the subjects the longest time. The test
showed that there were significant time differences among the three
drawings ($p = 0.039$). Pairwise comparisons found that the
differences between f1 and f3, f2 and f3 were statistically
significant ($p < 0.05$).

\subsubsection{Error Rate}

\begin{table}[h]
\begin{center}
\begin{tabular}{|l| r r r |r r r |}
\hline Drawing ID   & c1 & c2 & c3 & f1 & f2 & f3 \\
\hline\hline Node task & 0.00 & 6.25 & 25.00 & 0.00 & 0.00 & 0.00 \\
\hline Path task & 0.00 & 0.00 & 0.00 & 18.75 & 0.00 & 31.25 \\
\hline \small{Node+Path task} & 0.00 & 0.00 & 0.00 & 18.75 & 12.50 & 37.50 \\
\hline
\end{tabular}

\caption{Error rates (\%) for all the drawings and tasks}
\label{tbl:err}

\end{center}
\end{table}

The error rates for all the six drawings and the three tasks can be
seen in Table~\ref{tbl:err}. For the Node task, all the responses
were correct except for c2 and c3; the error rates were $6.25\%$ and
$25\%$ respectively. A visible inspection suggested that the angular
resolution in c2 and c3 was the worst among all the drawings. It is
reasonable to imagine that when the angular resolution is poor, the
task will be harder, thus causing more errors in counting the number
of edges that a node has. Although in this experiment, the bad
angular resolution was the result of making crossings, angular
resolution is not necessarily linked to crossings in practice.

For the path search tasks, regardless of whether the nodes were
highlighted or not, the subjects made no errors for Set 1 drawings.
This was not surprising, since there was only one path (therefore
the shortest path) between nodes 1 and 2. For Set 2 drawings, when
the nodes were highlighted, the highest error rate was made with f3
($31.25\%$), followed by f1 ($18.75\%$). No errors were made with
f2. When no nodes were highlighted, and following the same pattern,
the highest error rate was with f3 ($37.5\%$), followed by f1
($18.75\%$), then f2 ($12.5\%$).

\bigskip

\subsection{Eye Movements Video and Questionnaire Analysis}
\label{sec:video}

First, for the path search tasks with Set 1 drawings, it can be
clearly seen that the speed of eye movements was the fastest with
c1. With c2, the overall eye movements were still smooth but became
slower. Although some subjects claimed that they were not affected
by crossings here, the response time data did show that the subjects
responded significantly slower with c2 than with c1, as stated in
sub-section~\ref{sec:results}. With c3, eye movements were very
slow, and more significantly, on the edge connecting nodes 13 and 6
(see c3 in Table~\ref{tbl:drawings}). There were also quite a lot of
back-and-forth eye movements around the crossing points on that
edge. This indicted that the viewer was uncertain about which way to
go. Clearly, the low-angle crossings in c3 caused slow and extra eye
movements, which contributed to the longest response time.
Subjects\rq~ comments on crossings included: crossings \lq\lq force
me to focus harder\rq\rq, \lq\lq help to improve my
concentration\rq\rq; \lq\lq crossings affect me except at right
angles\rq\rq; \lq\lq I think edge crossings slowed me down\rq\rq;
\lq\lq crossings make graphs more complicated and confusing\rq\rq;
\lq\lq if the angle is small, you have to be careful when following
the edge to make sure you end up at the right node\rq\rq.

Secondly, for finding the shortest path in f1, most of the subjects
searched on the paths near the geometric path first. For example, 15
subjects for the Path task and 12 for the Node+Path task searched
the nearest path of 1-20-8-18-2 first. Some simply missed the
correct path of 1-11-15-2. Most of them detected the path either at
a later stage of search or just before pressing the button, as a
subject commented: \lq\lq I often found the shortest route
last\rq\rq.

The high error rate with f3 was surprising. The video inspection on
f3 revealed that the subjects spent most of their time on the left
part of the drawing, where there were more crossings. This also
happened with f2. In addition, most of the subjects found the
correct path (1-22-11-2) in f3 at the later stage. Six subjects
mentioned in the questionnaire that long edges had some influence
and commented: \lq\lq long edges need more time to reconfirm\rq\rq;
\lq\lq the shortest path of few long steps (edges) outside many
short steps is harder to see\rq\rq. However, intuitively long-edge
paths in sparse areas are visually more outstanding and should be
easier to detect~\cite{korner}, such as the shortest paths in f1 and
f3. This was not the case in this study and therefore needs further
examination.

Third, for the Node task, it seems that the subjects searched for
the most connected node randomly rather than systematically. This
probably was because the distribution of the nodes was unorganized
in the drawings. However, it is clear that the eye movement pattern
was as follows: eyes stayed around the node for a while counting the
edges, then moved straight from one node to the next. Subjects\rq~
eyes tended to start the task with nodes in dense areas first. All
the subjects claimed that for the Node task, crossings did not have
any influence on them. Some subjects preferred that edges are
attached to the same side of the node so that they can count the
edges at a glance, while others preferred that edges are evenly
distributed around the node. The significant difference in response
time between c1 and c3 for the Node task might be caused by the
difference in angular resolution between the two drawings.

\bigskip
\subsection{Comparison of Path Task and Node+Path Task}
\label{sec:comparision}

As mentioned in section~\ref{sec:tasks}, both Path task and
Node+Path task were included in the study to see whether and how
much the eyes can be distracted by crossings during the node
locating stage. The video analysis showed that the eyes of the
subjects appeared not to have been distracted by the crossings.
Their eye movement patterns were much similar to those for the Node
task, though their eyes moved faster since they only needed to check
the labels for the Node+Path task. This was supported by comments
from the subjects.

To compare the difference in response time, from
Table~\ref{tbl:time}, it can be seen that for each drawing, the
subjects spent more time for the Node+Path task than for the Path
task. This is normal since they needed extra time to locate the
nodes first. However, the statistical test revealed that the
differences in response time between the Path task and the Node+Path
task were marginally significant ($p=0.046$). The extra component of
finding nodes in the Node+Path task contributed a little in response
time.

With regard to the error rate (see Table~\ref{tbl:err}), for Set 2
drawings, the error rate remained unchanged with f1 for both tasks,
that is, $18.75\%$. The subjects made slightly more errors with f2
and f3 for the Node+Path task compared to the Path task; the error
rate increased from $0$ to $12.5\%$ with f2 and from $31.25\%$ to
$37.5\%$ with f3. This might be because, when no nodes were
highlighted, the shortest path became less visible and relatively
harder for the viewer to detect.

\bigskip
\bigskip

\section{Discussion}
\bigskip

\subsection{Layout Effects}
The results of this experiment indicated that the eye movements of
the Node task were largely independent of edge crossings. This is in
line with the finding that crossings are important only when path
tracing is involved~\cite{hu06, korner04}. \lq\lq After all, the
presence or absence of crossed lines does not constrain the position
of nodes in the graph\rq\rq~\cite{korner04}.

In this particular study, the extra component of locating nodes in
the Node+Path task only made marginal differences in response time,
compared to the Path task. This seemed unusual at first. However, it
is in fact reasonable since the human visual system is good at
searching for a target among similar distractions
(e.g,~\cite{korner04}). In addition, during node searching, the
subjects can get some idea about the paths between the nodes through
their peripheral vision. This in turn helped to reduce the time for
path searching, thus compensating for the time needed for locating
nodes.

It should be noted that the different findings of Ware et
al.~\cite{ware02} and Korner et al.~\cite{korner} about the effects
of crossings, should not be understood as a contradiction, but
rather as additional evidence suggesting that the effects of
crossings differ according to the situation. The task used
in~\cite{korner} involved a reasoning process about ordered sets,
while the task in~\cite{ware02} was a simple path search task for
abstract graphs.

With respect to the effects of crossing angles, the subjects spent a
significantly longer time with c3 than with c2. Although we are
unable to attribute this time difference exclusively to the small
crossing angles in c3 (c3 also has a larger number of crossings),
the eye movement data and user comments had made it clear that the
extra back-and-forth eye movements at and around the crossing points
were caused by the sharp crossing angles.

In the previous study~\cite{hu05}, in which the graphs were small
and the target shortest path was short (three at most), the impact
of crossings was not so apparent. However, in this study, where a
larger graph was used for c2 and the length of the shortest path was
six, both eye movement data and quantitative data clearly indicted
that crossings affected graph reading behavior and task performance
negatively. We may conjecture that when a graph is larger and the
searching path is longer, the impact of crossings can build up and
become significant, even if edges cross at nearly 90 degrees.

Although the conjecture needs confirmation, an immediate implication
for graph drawing is: when the graph to be drawn is small, we may
need to determine whether it is worthwhile removing crossings; when
the graph is large, it may be better to minimize the number of
crossings first. On the other hand, when crossings are to be present
in the final drawing, crossing angles should be increased to nearly
90 degrees if possible.

The effects of crossings on eye movements can be summarized as
follows: when edges cross at nearly 90 degrees, eye movements may be
slightly slowed down, but still smooth. When edges cross at small
angles, crossings cause confusion, slowing down and triggering extra
eye movements.

Geometric-path tendency is a graph reading convention; it seems
irrelevant to layout effects at first glance. However, it is highly
related to crossings. In many cases, it is crossings that cause
confusion, making all the paths between two nodes, and branches
along these paths, unforeseeable. Due to the geometric-path
tendency, human eyes can easily slip into the edges that are close
to the geometric path but not part of the target path.

\begin{figure}[t]

\centering
\includegraphics[height=1.7in]{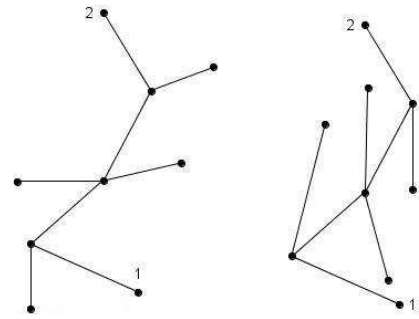}
\small \caption{Geometric-path tendency suggests that the path
between nodes 1 and 2 in the left drawing should be easier to detect
than that in the right. } \label{fig:tendency}
\end{figure}

Further, the relevance of geometric-path tendency to layout effects
goes beyond crossings. As illustrated in Figure~\ref{fig:tendency},
by changing the positions of the unlabeled nodes in the left
drawing, some edges are made to go toward the labeled nodes, as
shown in the right drawing. As a result of this, the path between
nodes 1 and 2 may become harder to follow. Therefore, understanding
how people read graphs in general and the geometric-path tendency in
particular, should assist us in better assessing the effectiveness
of a particular layout. More importantly, geometric-path tendency is
independent of specific graph layouts. If this tendency can be
extended for general path search tasks in future research, it is
expected that geometric-path tendency will be a more reliable factor
in predicting performance, compared to layout features such as
crossings.

\bigskip

\subsection{Eye Tracking in Graph Evaluation}

Eye tracking has been successfully used in psychology as well as in
HCI for many years~\cite{rayner98}. Very few studies, however, are
available in graph evaluation (understanding graphs that are drawn
as node-link diagrams)~\cite{korner}.

This study demonstrated the promising usefulness of eye tracking in
this area. Eye movement data offers additional insights into how
tasks are actually carried out, which is otherwise difficult to
obtain with traditional performance measures alone. In other words,
the eye tacking technology can be used to approve and refine the
theories developed with performance measures. In particular, since
graph evaluations are all about how people visually process
information from node-link diagrams, the use of eye tracking should
hold great promise for future research.

\bigskip
\bigskip

\section{Future Work}

Given the purposes of the study and the settings of the experiment,
the insights obtained in this study are far from conclusive. The
quantitative evidence from further fine-tuned experiments is needed
to verify them. In particular, a larger number of subjects and
graphs should be employed in a more natural environment. When eye
tracking is involved, standard ocular metrics~\cite{gold} such as
number of fixations and mean fixation duration should be included in
eye movement analysis. This work is currently ongoing. More specific
questions to be tested are:

\begin{enumerate}

\item Crossing angle has a significant effect on task performance. The
performance becomes worse when the angle decreases.

\item  Validity of geometric-path tendency and its impact on performance.

\end{enumerate}


\end{document}